\documentclass[reprint, aps,prd, preprintnumbers,groupedaddress,nofootinbib]{revtex4-1}
\usepackage{graphicx}
\graphicspath{{./plots/}}
\usepackage{latexsym}
\usepackage{amsfonts}
\usepackage{amssymb}
\usepackage{amsmath}
\usepackage{slashed}
\usepackage{feynmp}
\usepackage{hyperref}
\usepackage{url}
\usepackage{color}
\usepackage{cancel}
\usepackage[normalem]{ulem}
\usepackage{multirow,array}
\usepackage{dcolumn}
\usepackage{bm}
\usepackage[normalem]{ulem}

\newcommand\one{\leavevmode\hbox{\small1\normalsize\kern-.33em1}}

\def\slashchar#1{\setbox0=\hbox{$#1$}           
   \dimen0=\wd0                                 
   \setbox1=\hbox{/} \dimen1=\wd1               
   \ifdim\dimen0>\dimen1                        
      \rlap{\hbox to \dimen0{\hfil/\hfil}}      
      #1                                        
   \else                                        
      \rlap{\hbox to \dimen1{\hfil$#1$\hfil}}   
      /                                         
   \fi}

\newcommand{\be}{\begin{eqnarray*}}
\newcommand{\ee}{\end{eqnarray*}}

\newcommand{\bee}{\begin{eqnarray}}
\newcommand{\eee}{\end{eqnarray}}
\newcommand{\beeq}{\begin{equation}}
\newcommand{\eeeq}{\end{equation}}

\begin{document}

\title{Off-shell Higgs Couplings in  $H^*\to ZZ\to \ell\ell\nu\nu$}

\author{Dorival Gon\c{c}alves} 
\affiliation{Department of Physics, Oklahoma State University, Stillwater, OK, 74078, USA}
\author{Tao Han} 
\author{Sze Ching Iris Leung} 
\author{Han Qin} 

\affiliation{PITT PACC, Department of Physics and Astronomy, University of Pittsburgh, 3941 O'Hara St., Pittsburgh, PA 15260, USA}

\preprint{OSU-HEP-20-13,~~~ PITT-PACC-2012}

\begin{abstract}

We explore the new physics reach for the off-shell Higgs boson measurement in the ${pp \to H^* \rightarrow Z(\ell^{+}\ell^{-})Z(\nu\bar{\nu})}$ 
channel at the high-luminosity LHC. The new physics sensitivity is parametrized  in terms of the Higgs boson width, effective field theory framework, 
and a non-local Higgs-top coupling form factor.  Adopting  Machine-learning techniques, we demonstrate that the combination of a  large signal rate  
and a precise phenomenological probe for the process energy scale, due to the transverse $ZZ$ mass, leads to significant sensitivities beyond the 
existing results in the literature for the new physics scenarios considered.

\end{abstract}

\pacs{}
\maketitle

\section{Introduction}

After the Higgs boson discovery at the Large Hadron Collider (LHC)~\cite{Higgs:1964ia,PhysRevLett.13.508,PhysRevLett.13.321,Aad_2012,Chatrchyan_2012}, 
the study of the Higgs properties has been one of the top priorities in searching for new physics beyond the Standard Model (BSM). Indeed, the Higgs boson is a 
unique class in the SM particle spectrum and is most mysterious in many aspects. The puzzles associated with the Higgs boson include the mass hierarchy between
 the unprotected electroweak (EW) scale ($v$) and the Planck scale ($M_{PL}$), the neutrino mass generation, the possible connection to dark matter, the nature 
 of the electroweak phase transition in the early universe, to name a few. Precision studies of the Higgs boson properties can be sensitive to new physics at a higher 
 scale. Parametrically, new physics at a scale $\Lambda$ may result in the effects of the order $v^2/\Lambda^2$. 

So far, the measurements at the LHC based on the Higgs signal strength are in full agreement with the SM predictions. However, these measurements mostly focus on
the on-shell Higgs boson production, exploring the Higgs properties at low energy scales of the order $v$. It has been argued that if we explore the Higgs physics at a 
higher scale $Q$, the sensitivity can be enhanced as $Q^2/\Lambda^2$. A particularly interesting option is to examine the Higgs sector across different energy 
scales, using the sizable off-shell Higgs boson rates at the LHC~\cite{Kauer:2012hd,Caola:2013yja,Campbell:2013una,Aaboud:2018puo,Sirunyan:2019twz}.
While the off-shell Higgs new physics sensitivity is typically derived at the LHC with the $H^* \to ZZ \to 4\ell$ channel~\cite{Gainer:2014hha,Cacciapaglia:2014rla,Azatov:2014jga,Englert:2014aca,Buschmann:2014sia,Corbett:2015ksa,Goncalves:2017iub,Goncalves:2018pkt}, we demonstrate in this work that 
the extension to the channel $ZZ \to \ell \ell \nu \nu$ can significantly contribute to the potential discoveries. This channel provides two key ingredients to probe the
high energy regime with enough statistics despite of the presence of two missing neutrinos in the final state. First, it displays a larger event rate by a factor of six than
the four charged lepton channel. Second, the transverse mass for the $ZZ$ system sets the physical scale $Q^2$ and results in a precise phenomenological probe
to the underlying physics.

In this paper, we extend the existing studies and carry out comprehensive analyses for an off-shell channel in the Higgs decay 
\begin{equation}
pp \to H^* \to ZZ \to \ell^+ \ell^-\ \nu \bar \nu ,
\label{eq:process}
\end{equation}
where $\ell=e,\mu$ and $\nu=\nu_e,\nu_\mu,\nu_\tau$.
Because of the rather clean decay modes, we focus on the leading production channel of the Higgs boson via the gluon fusion. First, we 
phenomenologically explore a theoretical scenario with additional unobserved Higgs decay channels leading to an increase in the Higgs boson width, 
$\Gamma_H/\Gamma_H^{SM}>1$. The distinctive dependence for the on-shell and off-shell cross-sections with the Higgs boson width foster the conditions for a precise 
measurement for this key ingredient of the Higgs sector. We adopt the Machine-learning techniques in the form of Boosted Decision Tree (BDT) to enhance the signal sensitivity.
This analysis sets the stage for our followup explorations. Second, we study the effective field theory framework, taking advantage of the characteristic energy-dependence
 from some of the operators. Finally, we address a more general hypothesis that features a non-local momentum-dependent Higgs-top interaction~\cite{Goncalves:2018pkt},
namely, a form factor, that generically represents the composite substructure. Overall, the purpose of this paper is to highlight the complementarity across a multitude of 
frameworks~\cite{Englert:2014aca,Azatov:2014jga,Englert:2014ffa,Buschmann:2014sia,Corbett:2015ksa,Goncalves:2017iub,Goncalves:2018pkt} via the promising process at 
the LHC $H^*\rightarrow Z(\ell\ell)Z(\nu\nu)$, from models that predict invisible Higgs decays, passing by the effective field theory, and a  non-local form-factor scenario. 
Our results demonstrate significant sensitivities at the High-Luminosity LHC (HL-LHC) to the new physics scenarios considered here beyond the existing literature.

The rest of the paper is organized as follows. In Sec.~\ref{sec:width}, we derive the Higgs width limit at HL-LHC. Next, in Sec.~\ref{sec:eft}, we study the new  physics sensitivity
within effective field theory framework. In Sec.~\ref{sec:form-factor}, we scrutinize the effects of a non-local Higgs-top form-factor. Finally,  we  present a summary in Sec.~\ref{sec:summary}.

\section{Higgs Boson Width}
\label{sec:width}

\begin{figure}[tb]
\centering
\includegraphics[width=0.5\textwidth]{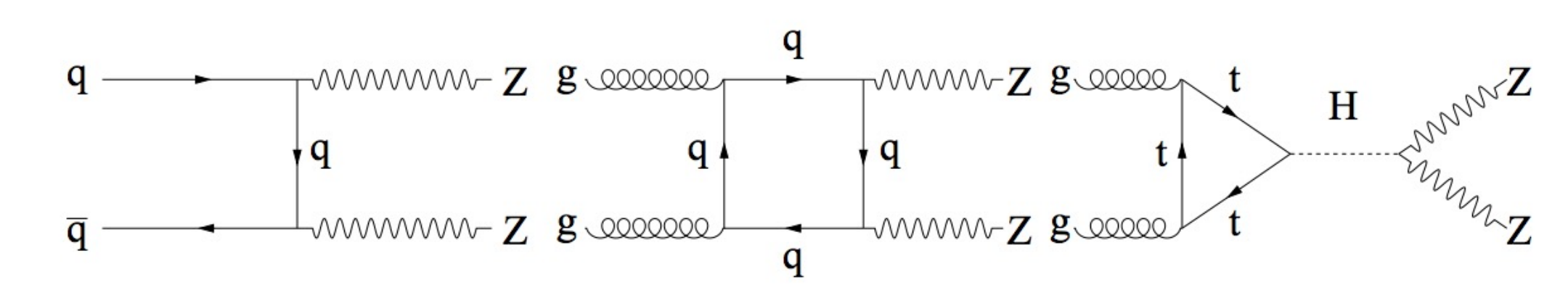}
\caption{Representative Feynman diagrams for the DY $q\bar{q}\rightarrow ZZ$ (left), GF $gg\rightarrow ZZ$ continuum (center), and $s$-channel Higgs signal 
$gg\rightarrow H^{*}\rightarrow ZZ$ (right).
\label{fig:feyn}}
\end{figure}

 \begin{figure*}[tb]
\centering
\includegraphics[width=0.34\textwidth]{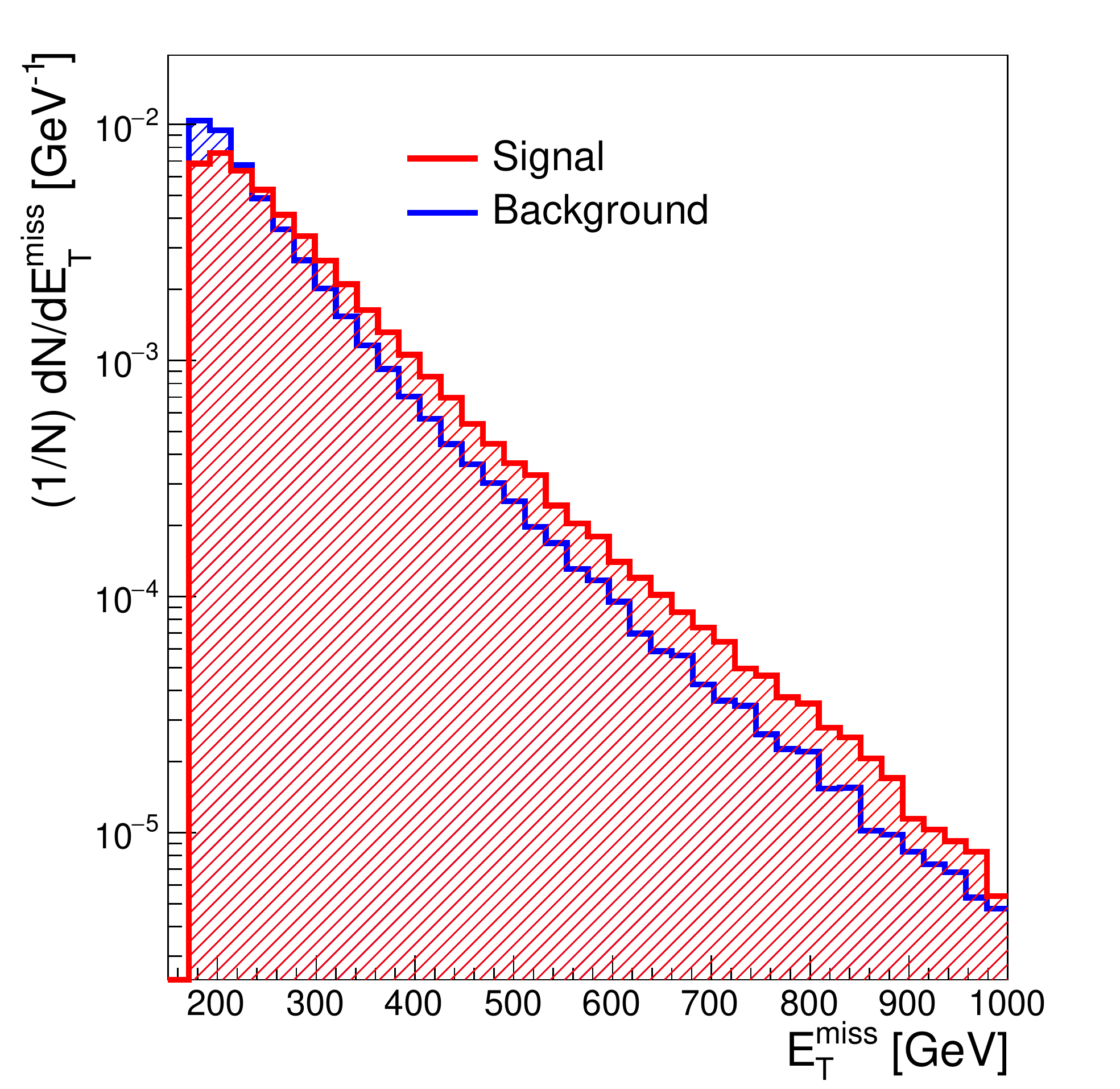}\hspace{-0.3cm}
\includegraphics[width=0.34\textwidth]{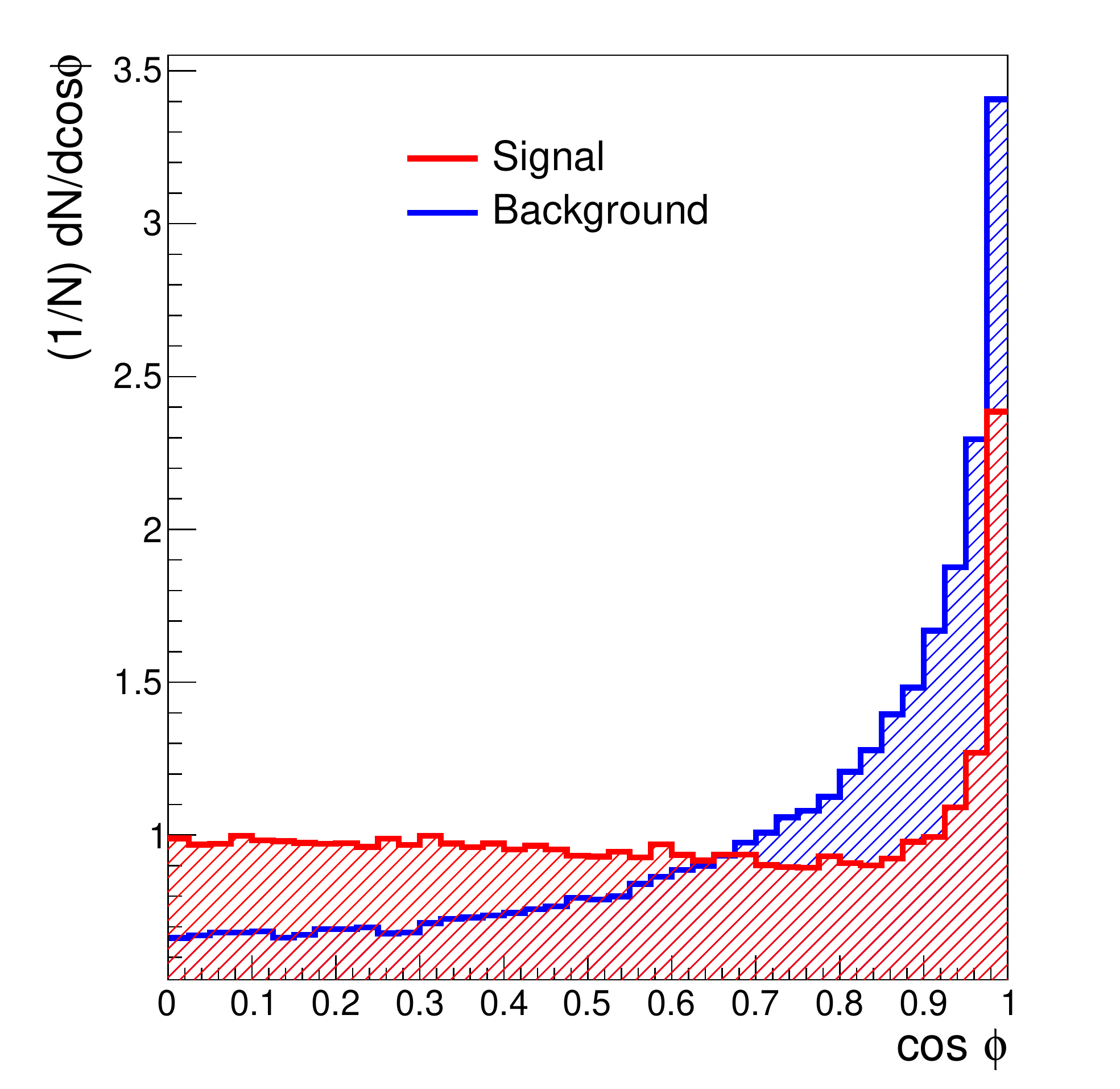}\hspace{-0.3cm}
\includegraphics[width=0.34\textwidth]{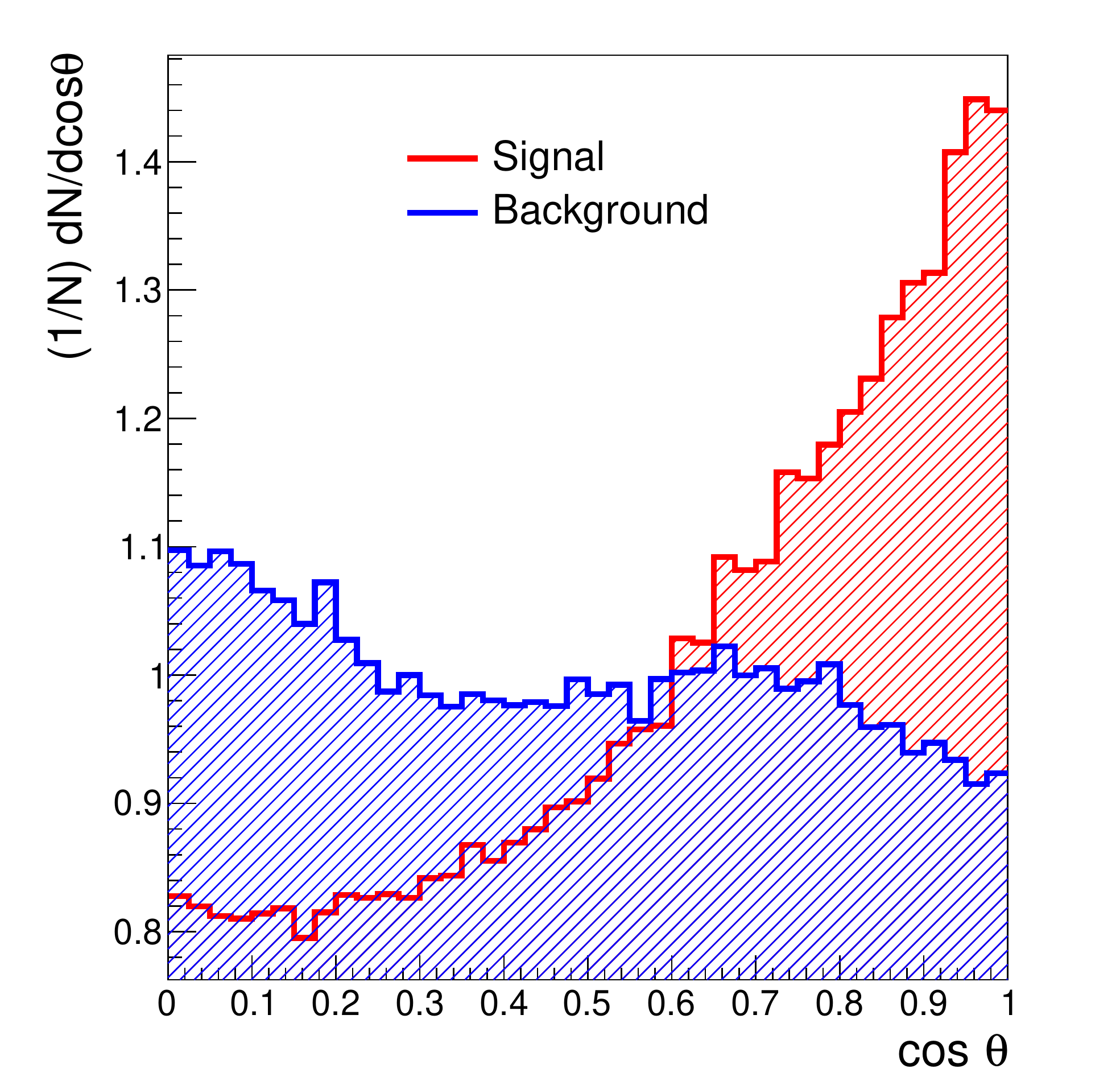}
\caption{Normalized distributions for the missing transverse momentum $E_T^{\mathrm{miss}}$ (left panel), azimuthal $\phi$ (central panel) and polar  
$\theta$ angles (right panel)  of the charged lepton $\ell^- $ in the $Z$ boson rest frame. 
\label{fig:bdtobs}}
\end{figure*}

The combination of on-shell and off-shell Higgs boson rates addresses one of the major shortcomings of the LHC, namely the Higgs boson width 
measurement~\cite{Kauer:2012hd,Caola:2013yja}. This method breaks the degeneracy present on the on-shell Higgs coupling studies
\begin{equation}
\sigma_{i\rightarrow H \rightarrow f}^{\text{on-shell}}\propto \frac{g_i^2(m_H)g_f^2(m_H)}{\Gamma_H}\,,
\end{equation}
where the total on-shell rate can be kept constant under the transformation $g_{i,f}(m_H)\rightarrow \xi g_{i,f}(m_H)$ with 
$\Gamma_H\rightarrow \xi^4\Gamma_H$. The  off-shell Higgs rate,  due to a sub-leading dependence on the Higgs boson width $\Gamma_H$
\begin{equation}
\sigma_{i\rightarrow H^* \rightarrow f}^{\text{off-shell}}\propto g_i^2(\sqrt{\hat{s}})g_f^2(\sqrt{\hat{s}})\,,
\end{equation}
breaks this degeneracy, where $\sqrt{\hat s}$ is the partonic c.m.~energy that characterizes the scale of the off-shell Higgs. In particular, if the 
new physics effects result in the same coupling modifiers at both kinematical regimes~\cite{Englert:2014aca,Azatov:2014jga,Buschmann:2014sia,Corbett:2015ksa},
 the relative measurement of the on-shell and off-shell signal strengths can uncover the Higgs boson width, 
$\mu_{\text{off-shell}}/\mu_{\text{on-shell}}=\Gamma_H/\Gamma_H^{SM}$.

In this section, we derive a projection for the Higgs boson width measurement at the ${\sqrt{s}=14}$~TeV high-luminosity LHC, exploring the 
$ZZ\rightarrow 2\ell2\nu$ final state. We consider the signal channel as in Eq.~(\ref{eq:process}). The signal is characterized by two same-flavor opposite sign  
leptons, $\ell=e$ or $\mu$, which reconstruct a $Z$ boson and recoil against a large  missing  transverse momentum from $Z\to \nu\bar \nu$. The major backgrounds for this search are
the Drell-Yan (DY) processes $q\bar{q}\rightarrow ZZ, ZW$ and gluon fusion (GF) $gg\rightarrow ZZ$ process, see Fig.~\ref{fig:feyn} for a sample of the Feynman 
diagrams.  While the Drell-Yan component displays the largest rate, the gluon fusion box diagrams interfere with the Higgs signal, resulting in important contributions 
mostly at the off-shell Higgs regime~\cite{Kauer:2012hd}. 

In our calculations, the signal and background samples are generated with  MadGraph5\textunderscore aMC@NLO \cite{Alwall:2014hca,mg5loop}. The Drell-Yan 
background is generated at the NLO with the \textsc{MC@NLO} algorithm~\cite{Frixione:2002ik}.  Higher order QCD effects to the loop-induced gluon 
fusion component  are included via a universal $K$-factor~\cite{Campbell:2013una,Bonvini:2013jha}. Spin correlation effects for the $Z$ and $W$ bosons 
decays are obtained in our simulations with the {\sc MadSpin} package~\cite{Artoisenet:2012st}. The renormalization and factorization scales are set by the invariant 
mass of the gauge boson pair $Q= m_{VV}/2$, using the PDF set {\sc nn23nlo}~\cite{Ball:2013hta}. Hadronization and underlying event effects are simulated with 
\textsc{Pythia8}~\cite{Sjostrand:2014zea}, and detector effects are accounted for with the \textsc{Delphes3} package~\cite{Ovyn:2009tx}.

We start our analysis with some basic lepton selections. We require two same-flavor and opposite sign  leptons with $|\eta_\ell|<2.5$ and $p_{T\ell}>10$~GeV
in the invariant mass window $76~\text{GeV}<m_{\ell\ell}<106$~GeV. To suppress the SM backgrounds, it is required  large missing energy selection 
$E_{T}^{\text{miss}} > 175$~GeV and a minimum transverse mass for the $ZZ$ system ${m_T^{ZZ}>250}$~GeV, defined as 
\begin{widetext}
\begin{equation}
m_T ^{ZZ}= \sqrt{\left(\sqrt{m_Z^2 + p_{T(\ell\ell)}^2} + \sqrt{m_Z^2 + (E_T^{\mathrm{miss}})^2}\right)^2 - \left | \overrightarrow{p}_{TZ} + \overrightarrow{E}_T^{\mathrm{miss}}\right|^2} \, .
\end{equation}
\end{widetext}
The consistency of our event simulation and analysis setup is confirmed through a cross-check  with the ATLAS study in Ref.~\cite{Aaboud:2018puo}. 

To further control the large Drell-Yan background,  a Boosted Decision Tree (BDT) is implemented via the Toolkit for Multivariate Data Analysis with 
\textsc{ROOT (TMVA)}~\cite{tmva}. The BDT is trained to distinguish the full background events from the $s$-channel Higgs production. The variables 
used in the BDT are missing transverse energy, the momenta and rapidity for the leading and sub-leading leptons $(p_T^{\ell1}, \eta^{\ell1}, p_T^{\ell2},\eta^{\ell2})$,
the leading jet $(p_T^{j1},\eta^{j1})$, the separation between the two charged leptons $\Delta R_{\ell\ell}$, the azimuthal angle difference between the di-lepton system 
and the missing transverse energy $\Delta\phi(\vec{p}_{T}^{~\ell\ell},\vec{E}^{\mathrm{miss}}_T)$, and the scalar sum of jets and lepton transverse momenta $H_T$. Finally, 
we also include  the polar $\theta$   and azimuthal  $\phi$  angles of the charged lepton $\ell^- $ in the $Z$ rest frame~\cite{Goncalves:2018fvn,Goncalves:2018ptp}.
We choose the coordinate system for the $Z$ rest frame following Collins and Soper (Collins-Soper frame)~\cite{Collins:1977iv}. The signal and background distributions
for these observables are illustrated in Fig.~\ref{fig:bdtobs}.  We observe significant differences between the $s$-channel signal and background in the $(\theta,\phi)$ angle 
distributions. These kinematic features arise from the different $Z$ boson polarizations for the signal and background components at the large di-boson invariant mass 
 $m_{T}^{ZZ}$~\cite{Buschmann:2014sia,Goncalves:2017gzy}. Whereas the $s$-channel Higgs tends to have $Z_L$ dominance, the DY background is mostly $Z_T$ dominated.
 
\begin{figure}[b]
\centering
\includegraphics[width=0.5\textwidth]{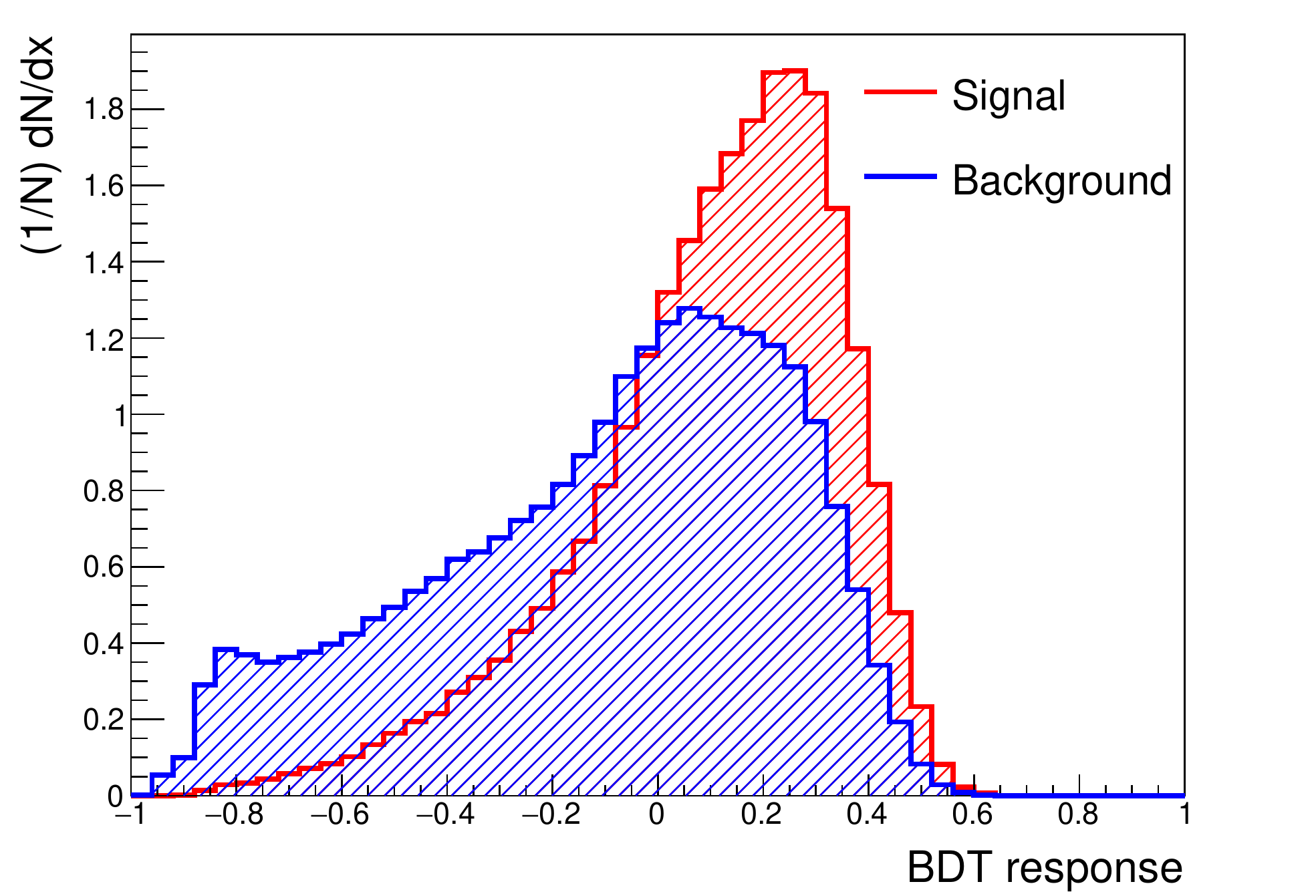}
\caption{BDT distribution for the $s$-channel Higgs signal (red) and background (blue). 
\label{fig:bdt-score}}
\end{figure}
 
We would like to illustrate the power of the implemented BDT analysis to separate the $s$-channel Higgs from the background contributions in Fig.~\ref{fig:bdt-score}.  
The BDT discriminator  is defined in the range $[-1,1]$. The events with discriminant close to $-1$ are classified as background-like and those close to 1 are signal-like.  
The optimal  BDT score selection has been performed with TMVA. To estimate the effectiveness of the BDT treatment, 
we note that one can reach $S/\sqrt{S+B}=5$ at an integrated luminosity of
$273~\text{fb}^{-1}$ with signal efficiency  88\% and background rejection of 34\%, by requiring ${\text{BDT}_{\text{response}}>-0.26}$. Now that we have tamed the  
dominant backgrounds ${q\bar{q}\rightarrow ZZ,ZW}$, we move on to the new physics sensitivity study. 

\begin{figure}[tb]
\centering
\includegraphics[scale = 0.47]{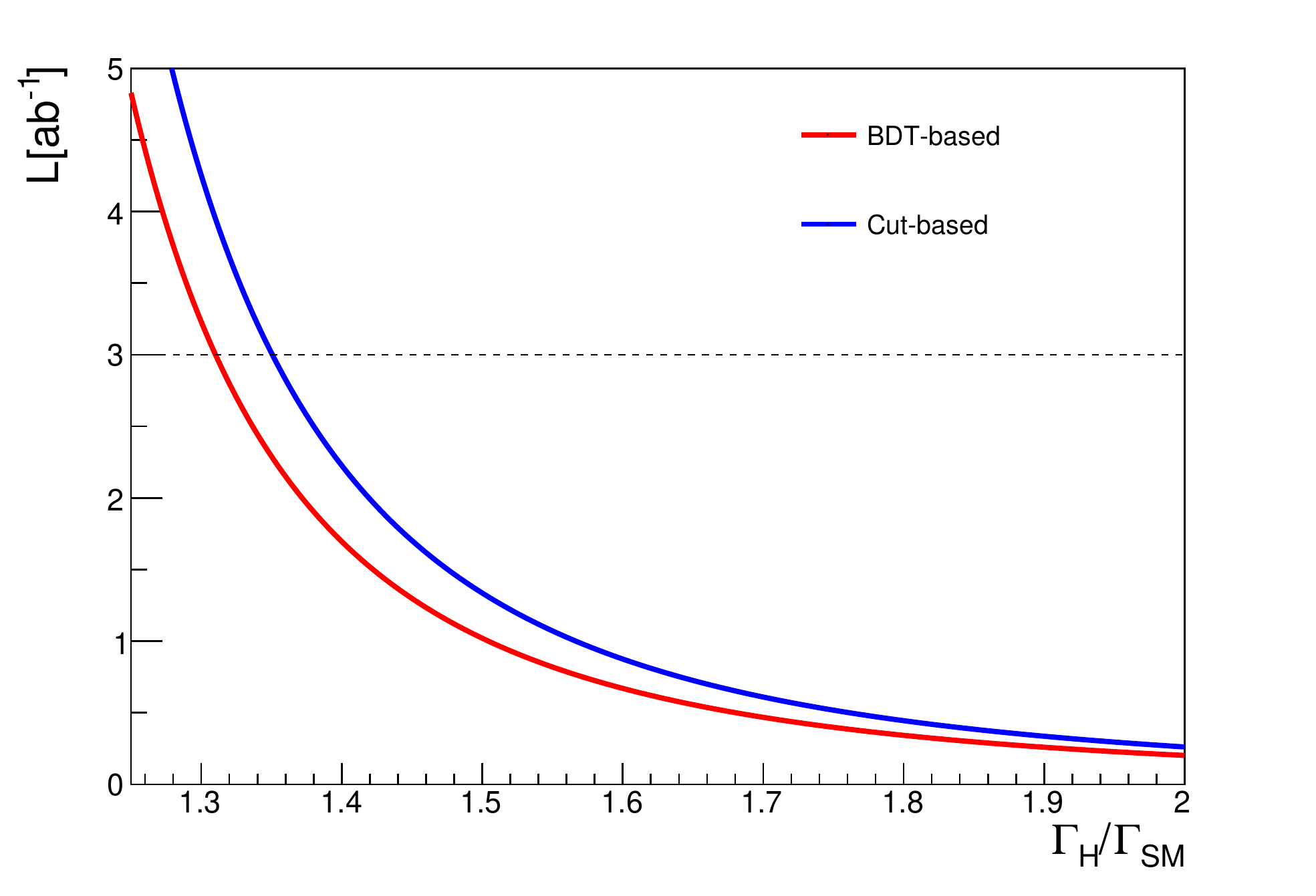}
\caption{95\% CL bound on the Higgs width $\Gamma_H/ \Gamma_H^{SM}$ as a function of the $\sqrt{s}=14$~TeV LHC luminosity. We display the results for
 the cut-based study (blue) and BDT-based analysis (red).}
\label{fig:width-bound}
\end{figure}

To maximize the sensitivity of the Higgs width measurement, we explore the most sensitive variable, 
$m_{T}^{ZZ}$ distribution, and perform a binned log-likelihood ratio analysis.
In Fig.~\ref{fig:width-bound}, we display the 95\%~CL on the Higgs width $\Gamma_H/ \Gamma_H^{SM}$ as a function of the $\sqrt{s}=14$~TeV LHC luminosity.
To infer the relevance of the multivariate analysis, that particularly explore the observables $(E_T^{\mathrm{miss}},\theta,\phi)$ depicted in Fig.~\ref{fig:bdtobs}, 
we display the results in two analysis scenarios: in blue we show the cut-based analysis and in red the results accounting for the BDT-based framework.  The significant 
sensitivity enhancement due to the BDT highlights the importance of accounting for the full kinematic dependence, including the $Z$-boson spin correlation effects.  
Whereas the  Higgs width can be constrained  to $\Gamma_H/ \Gamma_H^{SM}<1.35$  at 95\% CL level following the cut-based analysis,  
$\Gamma_H/ \Gamma_H^{SM}<1.31$ in the BDT-based study assuming $\mathcal{L}=3~\text{ab}^{-1}$ of data.  Hence, the BDT limits
result in an improvement of  $\mathcal{O}(5\%)$ on the final Higgs width sensitivity.  
These results are competitive to the HL-LHC
estimates for the four charged lepton final state derived by ATLAS and CMS, where the respective limits are $\Gamma_H/ \Gamma_H^{SM}<\mathcal{O}(1.3)$ 
and $\mathcal{O}(1.5)$ at 68\% CL~\cite{TheATLAScollaboration:2015stq,CMS:2018qgz}.

\section{Effective Field Theory}
\label{sec:eft}

\begin{figure}[tb]
\centering
\includegraphics[width=0.5\textwidth]{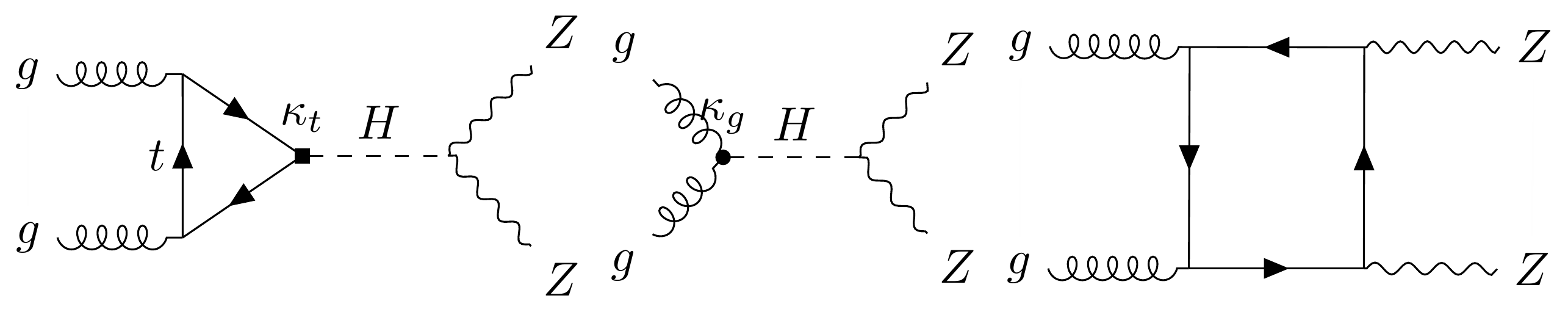}
\caption{Feynman diagrams for the GF $gg\rightarrow ZZ$  
process. The new physics effects from Eq.~(\ref{eq:eft2}) display deviations on the 
coefficients $\kappa_t$ and $\kappa_g$ from the SM point $(\kappa_t,\kappa_g)=(1,0)$.
\label{fig:feyn2}}
\end{figure}

The Effective Field Theory (EFT)  provides a consistent framework to parametrize beyond the SM effects in the presence of a mass gap
between the SM and new physics states. In this context, the new physics states can be integrated out and parametrized in terms of higher 
dimension operators~\cite{Appelquist:1974tg}. In this section we parametrize the new physics effects in terms of the EFT
framework~\cite{Buchmuller:1985jz,Grzadkowski:2010es}. Instead of performing a global coupling fit, we will focus on a relevant subset
of higher dimension operators that affect the Higgs production via gluon fusion. This will shed light on the new physics sensitivity for the 
off-shell $pp\rightarrow H^*\rightarrow Z(\ell\ell)Z(\nu\nu)$ channel. Our effective Lagrangian can be written as
\begin{alignat}{2}
\mathcal{L}\supset & c_g \frac{\alpha_s}{12\pi v^2}|\mathcal{H}|^2G_{\mu\nu}G^{\mu\nu}+
c_t\frac{y_t}{v^2}|\mathcal{H}|^2 \bar{Q}_L\tilde{\mathcal{H}}t_R+\text{h.c.} \,\,,
\label{eq:eft1}
\end{alignat}
where $\mathcal{H}$ is the SM Higgs doublet and $v=246$~GeV is the vacuum expectation value of the SM Higgs field. The couplings are normalized in such
a way for future convenience. If we wish to make connection with the new physics scale $\Lambda$, we would have the scaling as $c_g, c_t \sim v^2/\Lambda^2$. 
After electroweak symmetry breaking, Eq.~(\ref{eq:eft1}) renders into the following interaction terms with a single Higgs boson
\begin{alignat}{2}
\mathcal{L}\supset & \kappa_g \frac{\alpha_s}{12\pi v}H G_{\mu\nu}G^{\mu\nu}
-\kappa_t \frac{m_t}{v}H\left (\bar{t}_Rt_L+\text{h.c.}\right ) \,\,,
\label{eq:eft2}
\end{alignat}
where the coupling modifiers $\kappa_{g,t}$ and the Wilson coefficients $c_{g,t}$ are related by $\kappa_g=c_g$ and $\kappa_t=1-\text{Re}(c_t)$.
We depict in Fig.~\ref{fig:feyn2} the $gg\rightarrow ZZ$ Feynman diagrams that account for these new physics effects.

Whereas Eq.~(\ref{eq:eft1}) represents only a sub-set of high dimensional operators affecting the Higgs 
interactions~\cite{Buchmuller:1985jz,Grzadkowski:2010es}, we focus on it to highlight the effectiveness for the off-shell Higgs measurements to resolve a 
notorious degeneracy involving these terms. The gluon fusion Higgs production at low energy regime can be well approximated by the Higgs Low Energy 
Theorem~\cite{Shifman:1979eb,Kniehl:1995tn}, where the total Higgs production cross-section scales as ${\sigma_{\text{GF}}\propto |\kappa_t+\kappa_g|^2}$. 
Therefore,  low energy measurements, such as  on-shell and non-boosted Higgs production~\cite{Baur:1989cm,Harlander:2013oja,Banfi:2013yoa,Azatov:2013xha,Grojean:2013nya,Azatov:2014jga,
Buschmann:2014twa,Buschmann:2014sia,Azatov:2016xik}, are unable to resolve the ${|\kappa_t+\kappa_g|=\text{constant}}$ degeneracy.
While the combination between the $t\bar{t}H$ and gluon fusion Higgs production have the potential to break this blind direction~\cite{Plehn:2015cta}, 
we will illustrate that the Higgs production at the off-shell regime can also result into relevant contributions to resolve this degeneracy.

\begin{figure}[!b]
\centering
\includegraphics[width=0.5\textwidth]{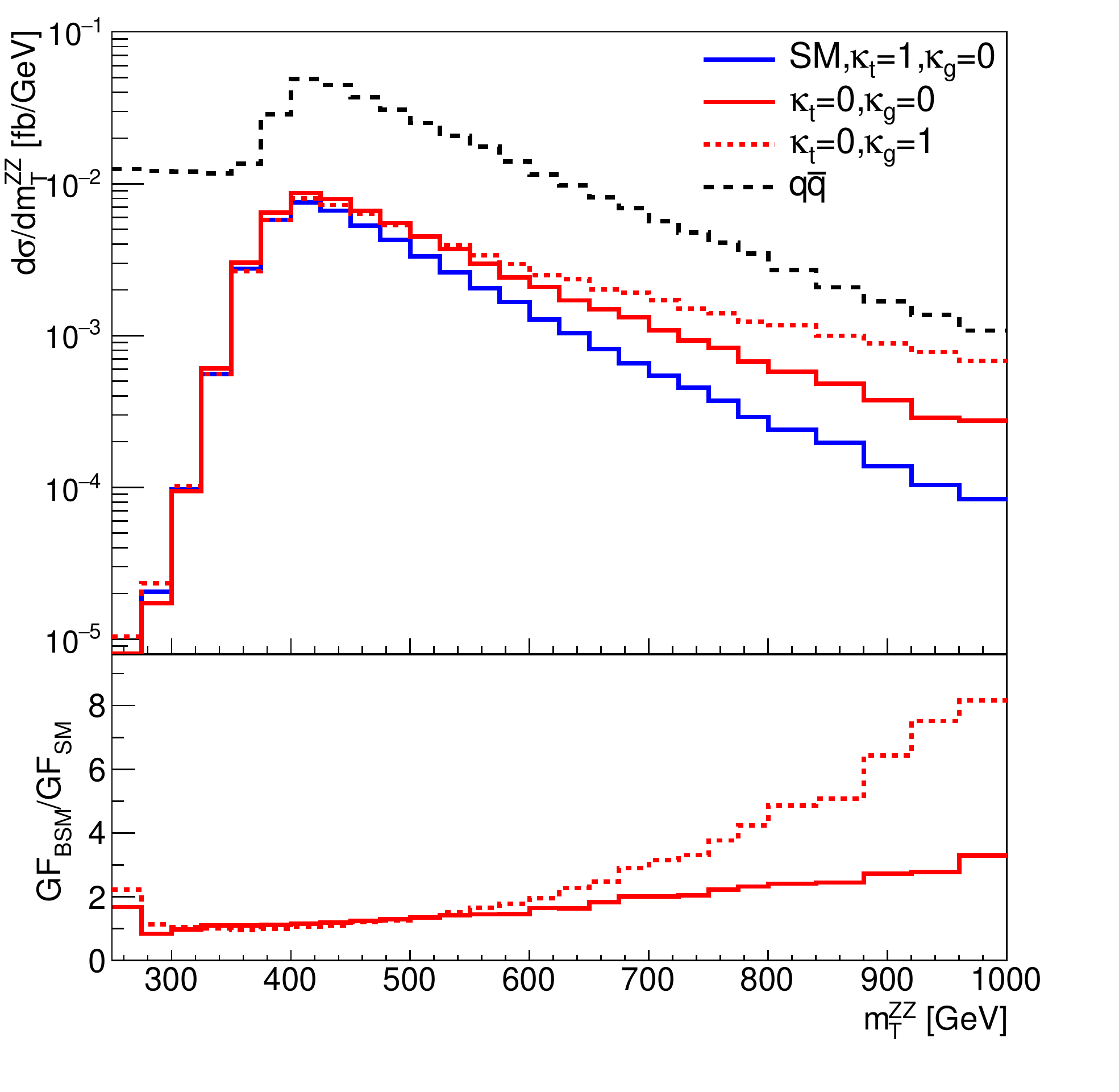}
\caption{Transverse mass distributions $m_T^{ZZ}$ for the DY and GF $Z(\ell\ell)Z(\nu\nu)$ processes. The new physics effects 
are parametrized by deviations from SM point $(\kappa_t,\kappa_g)=(1,0)$. We follow the benchmark analysis defined in Sec~\ref{sec:width}.
\label{fig:mzz}}
\end{figure}

Since the Higgs boson decays mostly to longitudinal gauge bosons at the high energy regime, it is enlightening  to inspect the signal
amplitude for the longitudinal components. The amplitudes associated to each contribution presented in Fig.~\ref{fig:feyn2} can be approximated 
at $m_{ZZ} \gg m_t,m_H,m_Z$  by~\cite{Glover:1988rg,Azatov:2014jga,Buschmann:2014sia}
\begin{alignat}{2}
\mathcal{M}_t^{++00} &\approx  + \frac{m_t^2}{2m_Z^2} \log^2\frac{m_{ZZ}^2}{m_t^2}\,, \notag \\
\mathcal{M}_g^{++00} &\approx  - \frac{m_{ZZ}^2}{2m_Z^2}\,,  \notag \\
\mathcal{M}_c^{++00} &\approx  - \frac{m_t^2}{2m_Z^2} \log^2\frac{m_{ZZ}^2}{m_t^2} \,. 
\label{eq:amp}
\end{alignat}
Two comments are in order. First, both the $s$-channel top loop  $\mathcal{M}_t$ and the continuum $\mathcal{M}_c$ 
amplitudes display logarithmic dependences on $m_{ZZ}/m_t$ at the far off-shell regime.  In the SM scenario the ultraviolet logarithm 
between these two amplitudes cancel, ensuring a proper high energy behavior when calculating the full amplitude. Second, it is worth 
noting the difference in sign between the $s$-channel contributions $\mathcal{M}_t$ and $\mathcal{M}_g$. This results into a destructive
interference  between  $\mathcal{M}_t$ and $\mathcal{M}_c$, contrasting to a constructive interference between  $\mathcal{M}_g$ and 
$\mathcal{M}_c$. In the following, we will explore these phenomenological effects pinning down the new physics sensitivity with a higher precision. 

\begin{figure}[tb]
\centering
\includegraphics[width=0.5\textwidth]{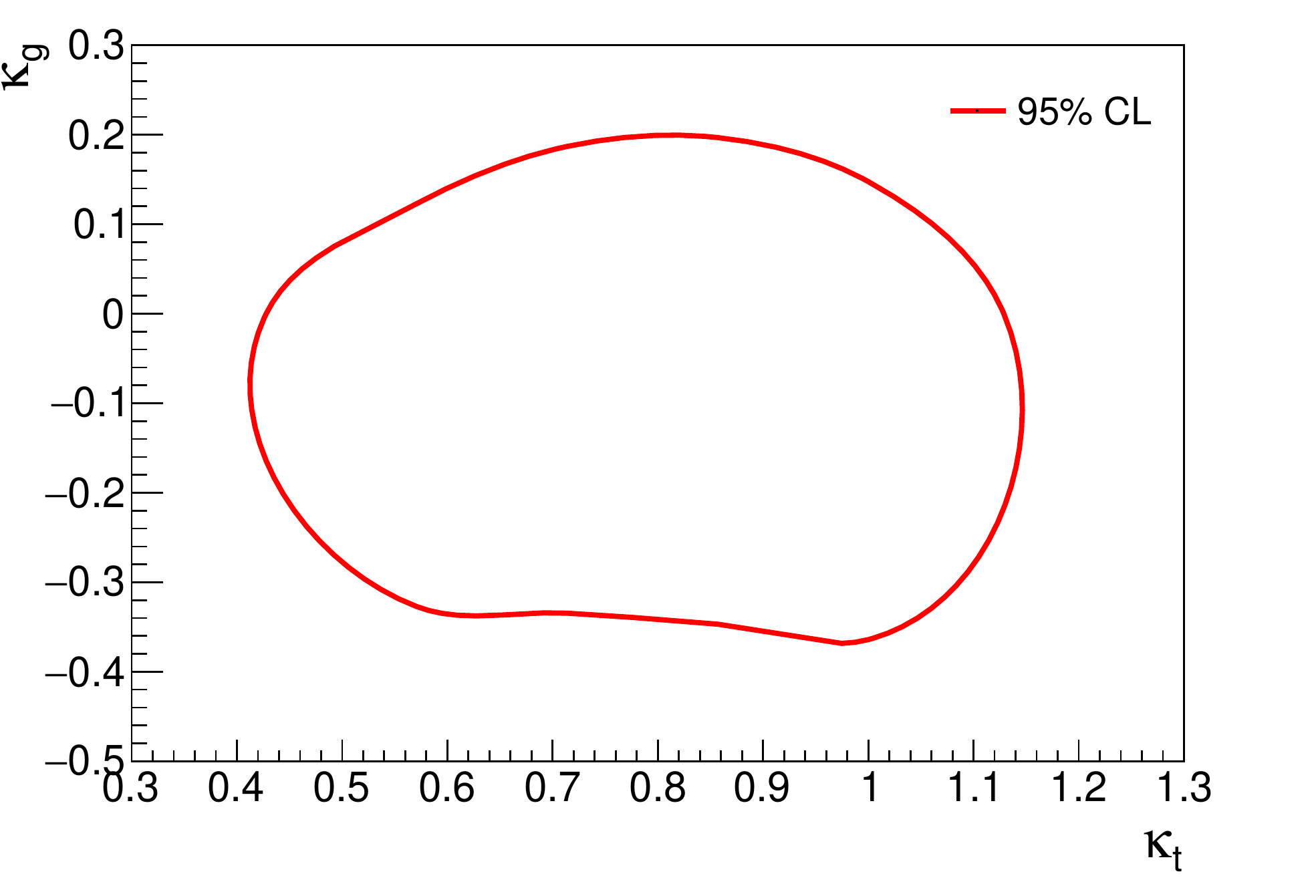}
\caption{95\% CL bound on the coupling modifiers $\kappa_t$ and $\kappa_g$  when accounting for the off-shell Higgs measurement 
in the $Z(\ell\ell)Z(\nu\nu)$  channel. We assume the 14~TeV LHC with $3~\text{ab}^{-1}$ of data.
\label{limit_kappa}}
\end{figure}

Exploiting the larger rate for $ZZ\rightarrow \ell\ell \nu\nu$ than that for $ZZ\rightarrow 4\ell$~\cite{Azatov:2014jga,Englert:2014aca,Buschmann:2014sia}, 
we explore the off-shell Higgs physics at the HL-LHC. To simulate the full loop-induced effects, we implemented Eq.~(\ref{eq:eft2}) into 
FeynRules/NLOCT~\cite{Alloul:2013bka,Degrande:2014vpa} through a new fermion state, and adjusting its parameters to match the low-energy  
Higgs interaction $HG_{\mu\nu}G^{\mu\nu}$~\cite{Shifman:1979eb,Kniehl:1995tn}. Feynman rules are exported to a Universal FeynRules Output 
(UFO)~\cite{Degrande_2012}  and the Monte Carlo event generation is performed with MadGraph5aMC@NLO~\cite{Alwall:2014hca}. 

In Fig.~\ref{fig:mzz}, we present the Drell-Yan (DY) and the gluon-fusion (GF) $m_{T}^{ZZ}$ distributions for different signal hypotheses. 
In the bottom panel, we display the ratio between the GF beyond the SM (BSM) scenarios with respect to the GF SM. In agreement with 
Eq.~(\ref{eq:amp}), we observe a suppression for the full process when accounting for the $s$-channel top loop contributions and an 
enhancement when including the new physics terms associated to $\mathcal{M}_g$ at high energies. 

We follow the benchmark analysis defined in Sec.~\ref{sec:width}. After the BDT study, the resulting events are used in a  binned
log-likelihood analysis with the $m_T^{ZZ}$ distribution. This approach explores the characteristic high energy behavior for the new
physics terms highlighted in Eq.~(\ref{eq:amp}) and illustrated in Fig.~\ref{fig:mzz}. We present in Fig.~\ref{limit_kappa} the resulting 
95\% CL sensitivity to the $(\kappa_t,\kappa_g)$ new physics parameters at the high-luminosity LHC. In particular, we observe that
the LHC can bound the top Yukawa within $\kappa_t\approx[0.4,1.1]$ at 95\% CL, using this single off-shell channel. The observed 
asymmetry in the limit, in respect to the SM point, arises from the large  and negative interference term between the $s$-channel and
the continuum amplitudes. The upper bound on $\kappa_t$ is complementary to the direct Yukawa measurement via 
$ttH$~\cite{Cepeda:2019klc}  and can  be further improved through a combination with the additional relevant off-shell Higgs final states. 
The results derived in this section are competitive to the CMS HL-LHC prediction that considers the boosted Higgs production combining the 
$H\to 4\ell$ and $H\to \gamma\gamma$ channels~\cite{CMS:2018qgz}. The CMS projection results into an upper bound on the top Yukawa of 
$\kappa_t\lesssim 1.2$ at 95\%~CL.

\section{Higgs-Top Form Factor}
\label{sec:form-factor}

\begin{figure}[b]
\centering
\includegraphics[width=0.5\textwidth]{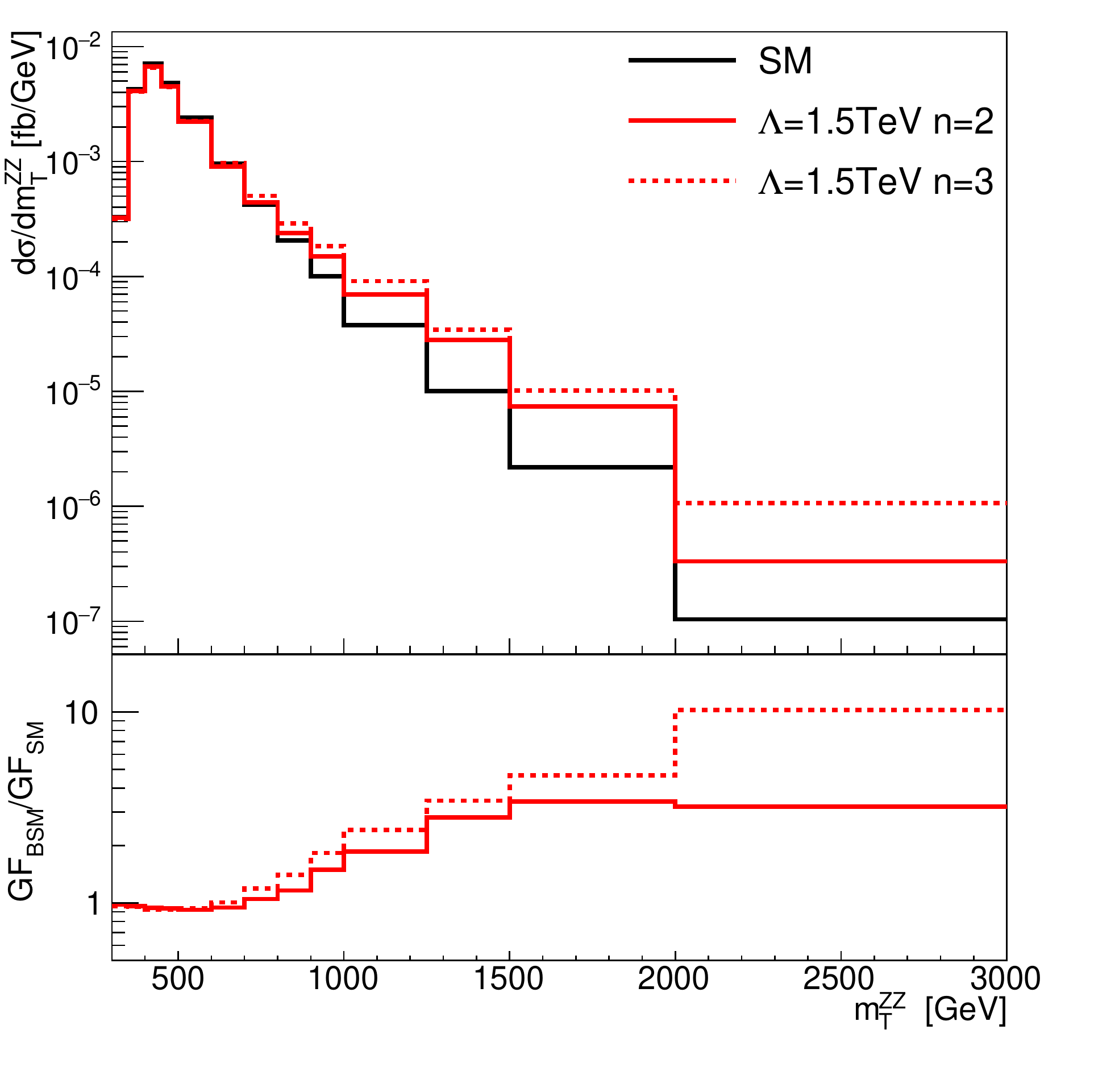}
\caption{Transverse mass distribution $m_T^{ZZ}$ for $ g g (  \rightarrow H^*  )  \rightarrow  Z(2\ell) Z(2\nu)  $ in the  Standard Model (black) 
and with a new physics form factor (red). We assume $n=2, 3$ and $\Lambda=1.5$~TeV for the form factor scenario.
\label{mzz_form_fac}}
\end{figure}

\begin{table*}[t]
\centering
\begin{tabular}{|c|c|c|c|}
\hline
&  $\Gamma_H/ \Gamma_H^{SM}$ & $\Lambda_{EFT}$ & $\Lambda_{Composite}^{n=2}$   \\
\hline
$H^{*}  \rightarrow Z Z \to \ell \ell\nu\nu$  &  1.31      &  0.8~TeV   &   1.5 TeV  \\
    \hline
$H^{*}  \rightarrow Z Z \to 4\ell$  &   1.3 (68\% CL)~\cite{TheATLAScollaboration:2015stq} &  0.55~TeV~\cite{CMS:2018qgz} & 0.8 TeV \cite{Goncalves:2018pkt}  \\
       \hline
\end{tabular}
\caption{Comparison of the sensitivity reaches between $H^{*}  \rightarrow Z Z \to \ell \ell \nu \nu$ in this study and $H^{*}  \rightarrow Z Z \to 4\ell$ in the literature 
as quoted. All results are presented at 95\% CL except for the Higgs width projection derived by ATLAS with 68\% CL~\cite{TheATLAScollaboration:2015stq}. 
We assume that the Wilson coefficient for the EFT framework is given by ${ c_{t} = v^2/\Lambda_{EFT}^2}$. Besides the $H\to 4\ell$ channel, Ref.~\cite{CMS:2018qgz} 
also accounts for the  $H\to \gamma\gamma$ final state with a boosted Higgs analysis.}
\label{tab:sum}
\end{table*}

The fact that the observed Higgs boson mass is much lighter than the Planck scale implies that there is an unnatural cancellation between the bare
 mass and the quantum corrections. Since the mass of the Higgs particle is not protected from quantum corrections, it is well-motivated to consider
 that it may not be fundamental, but composite in nature~\cite{Pomarol:2012qf,Panico:2011pw,Panico:2015jxa,Liu:2017dsz}. In such a scenario, the Higgs boson
is proposed as a bound state of a strongly interacting sector with a composite scale $\Lambda$. In addition, the top quark, which is the heaviest particle in the SM, 
can also be composite. In this case, the top Yukawa coupling will be modified by a momentum-dependent form factor at a scale $q^2$ close to or above the
 new physics scale $\Lambda^2$. It is challenging to find a general construction for such form factor without knowing the underlying dynamics. 
Here, we will adopt a phenomenological ansatz motivated by the nucleon form factor~\cite{Punjabi:2015bba}.  It is defined as
\begin{equation}
\label{eq:form-fac}
\Gamma(q^2/\Lambda^2) = \frac{1}{(1+q^2/\Lambda^2)^n}\,,
\end{equation}
where $q^2$ is the virtuality of the Higgs boson. For $n=2$, it is a dipole-form factor and corresponds to an exponential spacial distribution.
Building upon Ref.~\cite{Goncalves:2018pkt}, we study the impact of this form factor on
$ g g\rightarrow H^{*}  \rightarrow Z Z  $ process now with the complementary final state $ \ell^+ \ell^- \nu \overline{\nu}$.

\begin{figure}[b]
\centering
\includegraphics[width=0.5\textwidth]{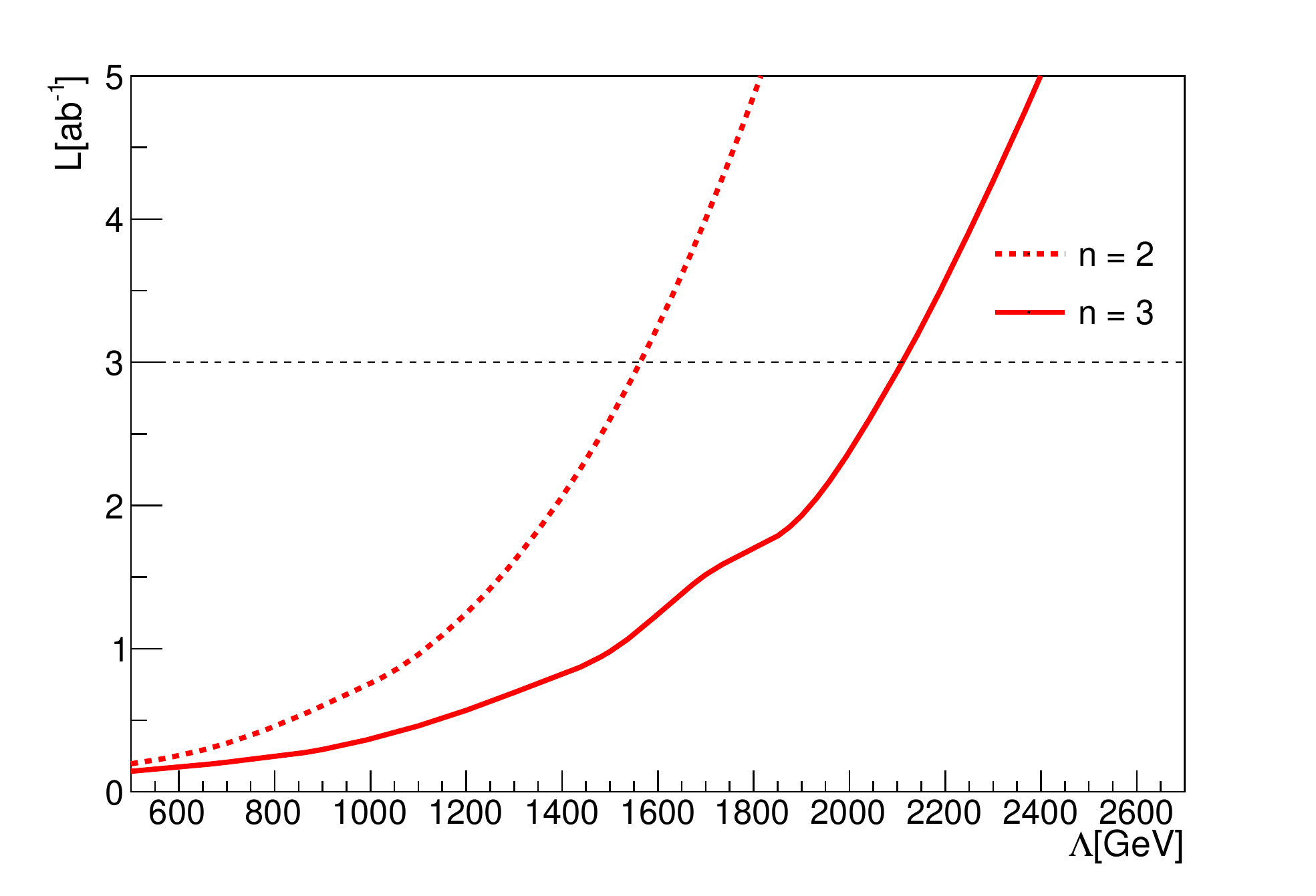}
\caption{95\% CL sensitivity on the new physics scale $\Lambda$
 as a function of the LHC luminosity. We assume the form factor in Eq.~(\ref{eq:form-fac})
 with $n=2$  (dashed line) and $n=3$ (solid line) at the 14~TeV LHC.
\label{bound_form_fac}}
\end{figure}

In Fig.~\ref{mzz_form_fac}, we illustrate the $m_{T}^{ZZ}$ distribution for the  full gluon fusion $ g g ( \rightarrow H^* )  \rightarrow Z Z  $ process. We show the  Standard Model  (black) and the form factor scenario (red). We assume $n=2$ or $3$ and $\Lambda=1.5$~TeV for the depicted form factor 
scenarios. The differences between Standard Model and form factor cases become larger when the energy scales are comparable or above $\Lambda$ due
to the suppression of destructive interference between Higgs signal and continuum background. Thus, we  perform  the same BDT procedure introduced in Sec.~\ref{sec:width} followed by a 
binned log-likelihood ratio test in the $m_T^{ZZ}$ distribution to fully explore this effect. In Fig.~\ref{bound_form_fac}, we display the sensitivity reach for the LHC in the Higgs-top form factor.
We observe that the LHC can bound  these new physics effects up to $\Lambda = 1.5$~TeV for $n = 2 $ and $\Lambda = 2.1$~TeV for $n=3$ at 95\% CL. The large event rate for the  $H^*\rightarrow ZZ\rightarrow \ell\ell\nu\nu$ signal results in a more precise probe to the ultraviolet regime than for the 
$H^*\rightarrow ZZ\rightarrow 4\ell$ channel, where the limits on the new physics scale are $\Lambda = 0.8$~TeV for $n = 2 $ and $\Lambda = 1.1$~TeV for $n=3$ at 95\% CL~\cite{Goncalves:2018pkt}.

\section{Summary}
\label{sec:summary}

We have systematically studied the off-shell Higgs production in the  $pp\rightarrow H^{*}\rightarrow Z(\ell\ell)Z(\nu\nu)$ channel
at the high-luminosity LHC. We showed that this signature is crucial to probe the Higgs couplings across different energy scales  potentially
shedding light on new physics at the ultraviolet regime. To illustrate its physics potential, we derived the LHC sensitivity to three BSM benchmark scenarios 
where the new physics effects are parametrized in terms of the Higgs boson width, the effective field theory framework, and a non-local Higgs-top coupling form factor. 

The combination of a large signal rate and a precise phenomenological probe for the process energy scale, due to the transverse $ZZ$ mass, 
renders strong limits for all considered BSM scenarios. A summary table and comparison with the existing results in the literature are provided in Table~\ref{tab:sum}. 
Adopting  Machine-learning techniques, we demonstrated in the form of BDT that the HL-LHC, with $\mathcal{L}=3~\text{ab}^{-1}$ of data, will display large  
sensitivity to the Higgs boson width, $\Gamma_H/ \Gamma_H^{SM}<1.31$. In addition, the characteristic high
energy behavior for the new physics terms within the EFT framework results in relevant bounds on the $(\kappa_t,\kappa_g)$ new physics 
parameters, resolving the low energy degeneracy in the gluon fusion Higgs production.  In particular, we observe that the LHC can bound 
the top Yukawa within $\kappa_t\approx[0.4,1.1]$ at 95\% CL. The upper bound on $\kappa_t$ is complementary to the direct Yukawa 
measurement via  $ttH$  and can  be further improved in conjunction with additional relevant off-shell Higgs channels.  Finally, when 
considering a more general hypothesis that features a non-local momentum-dependent Higgs-top interaction, we obtain that the HL-LHC is 
sensitive to new physics effects at large energies with $\Lambda = 1.5$~TeV for $n = 2 $ and $\Lambda = 2.1$~TeV for $n=3$ at 95\% CL.
We conclude that, utilizing the promising $H^*\to Z(\ell^+\ell^-)Z(\nu\bar \nu)$ channel at the HL-LHC and adopting the Machine-Learning techniques, the 
combination of a large signal rate and a precise phenomenological probe for the process energy scale renders improved sensitivities  beyond the existing literature,
 to all the three BSM scenarios considered in this work.

\begin{acknowledgments}
This work was supported by the U.S.~Department of Energy under grant No.~DE-FG02- 95ER40896 and by the PITT PACC.
DG was supported by the U.S.~Department of Energy under grant number DE-SC 0016013.
 \end{acknowledgments}

 \bibliography{reference}{}
 \bibliographystyle{unsrt}

\end{document}